\documentclass[prl,twocolumn,amsmath,amssymb]{revtex4}
\usepackage{graphicx}
\usepackage{dcolumn}
\usepackage{subfigure}
\usepackage{bm}
\usepackage{color}

\begin{document}

\title{Nambu-Goldstone modes and diffuse deformations in elastic shells}
\author{Christian D. Santangelo}
\email{csantang@physics.umass.edu}
\affiliation{
Department of Physics, University of Massachusetts, Amherst, MA  01003}
\date{\today}

\begin{abstract}
I consider the shape of a deformed elastic shell. Using the fact that the lowest-energy, small deformations are along infinitesimal isometries of the shell's mid-surface, I describe a class of weakly-stretching deformations for thin shells based on the Nambu-Goldstone modes associated with those isometries. The main result is an effective theory to describe the diffuse deformations of thin shells that incorporate stretching and bending energies. The theory recovers previous results for the propagation of a ``pinch'' on a cylinder. A cone, on the other hand, has two length scales governing the persistence of a pinch: one governing the relaxation of the pinch that scales with thickness as a $-1/2$ power, and one that scales with thickness above which deformations again become isometric. These lengths meet at a critical thickness below which low energy deformations again become nearly isometric.
\end{abstract}


\maketitle

As long ago as 1849, the Rev. Jellett pointed out that thin shells have a class of deformations that are privileged due to the large separation in scales between bending and stretching energy \cite{jellett}. These isometries -- deformations of the shell's mid-surface that do not stretch it -- were the basis upon which Rayleigh's theory of small vibrations in shells was built \cite{rayleigh, rayleigh2}, and were used more recently in a study of the rigidity of elastic torii \cite{audoly,audoly1}. In more modern applications, isometries appear in the guise of developable surfaces, characterizing all the forms that can be obtained by bending, but not stretching, a flat, elastic sheet \cite{docarmo}.

Under some types of forcing or constraints, however, an elastic sheet must stretch. The prototypical case is that of a crumpled sheet of paper \cite{witten}, and more specifically an intrinsically-flat elastic sheet pushed into a cup to form a d-cone \cite{dcone}. On a curved shell, matters are more complex. On the one hand, there are singular structures reminiscent of d-cones which sharpen with decreasing shell thickness \cite{vaziri, quilliet}. On the other, there are deformations of shells, such as ``pinching'' a straw, which result in very diffuse stretching over a large area \cite{maha}. Finally, coexistence between diffuse and sharp features have been observed in wrinkled sheets \cite{maha2, benny, pedro}.

This paper describes a generalization of shell isometries, analogous to Nambu-Goldstone modes in an ordered phase. I derive a ``universal'' effective theory for the small-amplitude, diffuse (as opposed to sharp) deformations of thin shells. My approach is similar to that of Lange and Newell for the post buckling behavior of a shell \cite{newell1, newell2}, but applies more generally. The theory agrees well with previous results on the persistence of a ``pinch'' in a cylinder \cite{maha}, and generalizes easily to more general shells. This is illustrated by considering the deformations of a cone, which qualitatively departs from those of a cylinder.

The mid-surface of a thin elastic shell is represented by a set of coordinates $(x_1,x_2)$ and functions $\mathbf{X}(x_1,x_2)$ describing its shape in space. For small strains, I follow standard practice in writing the energy as a sum of two terms, $E=E_{st} + E_{b}$ \cite{dias}. The first term, due to in-plane stretching, is
\begin{equation}\label{eq:stretchingenergy}
E_{st} = \frac{t}{8} \int d^2x~ \sqrt{\bar{g}} A^{i j k l} \gamma_{i j} \gamma_{k l},
\end{equation}
where $t$ is the thickness, $\bar{g}_{i j}$ is the reference metric, $\bar{g}$ its determinant, and $A^{i j k l} = \lambda \bar{g}^{i j} \bar{g}^{k l} + 2 \mu \bar{g}^{i k} \bar{g}^{j l}$ in terms of the two Lam\`e constants. Alternatively, $\lambda = Y/(1-\nu^2)$ and $2 \mu = Y/(1+\nu)$, where $Y$ is the Young's modulus and $\nu$ the Poisson ratio. The strain is the difference between realized and reference metrics, $\gamma_{i j} = g_{i j} - \bar{g}_{i j} = \partial_i \mathbf{X} \cdot \partial_j \mathbf{X} - \bar{g}_{i j}$. The second term in the energy, due to the bending of the sheet, is
\begin{equation}\label{eq:bendingenergy}
E_{b} = \frac{t^3}{24} \int d^2x~ \sqrt{\bar{g}} A^{i j k l} [h_{i j}-\bar{h}_{i j}] [h_{k l}-\bar{h}_{k l}],
\end{equation}
where $h_{i j} = \mathbf{N} \cdot \partial_i \partial_j \mathbf{X}$ is the second fundamental form, $\mathbf{N}$ is the unit normal of the mid-surface, and $\bar{h}_{i j}$ is a reference bending form. When the reference metric can be realized as a surface immersed in three dimensions, a free elastic sheet will expel in-plane strain as $t \rightarrow 0$ \cite{lewickashell}.

\begin{figure}[b]
\includegraphics[width=3in]{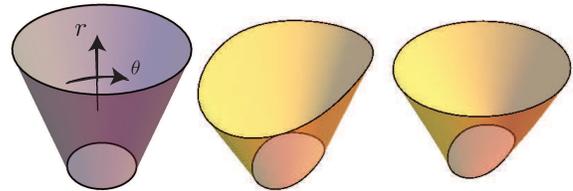}
\caption{(left) The coordinate $r$ is the arc length on the undeformed cone and $\theta$ is the azimuthal angle. (middle) A cone pinched at the inside edge along an infinitesimal isometry. (right) A cone pinched at the inside edge along an isometry that decays along a Nambu-Goldstone mode.}
\label{fig:cone}
\end{figure}

The softest deformations of a shell are the isometries -- deformations that change the shape of the mid-surface while leaving the metric invariant \cite{jellett}. More precisely, I consider a shell with an $N$-parameter family of isometries parameterized by $\epsilon_I$, where $I$ ranges from $1$ to $N$. An $n^{th}$ order isometry requires $\partial_\epsilon^k |_{\epsilon=0} (\partial_i \mathbf{X} \cdot \partial_j \mathbf{X}) = 0$ for any $k \le n$. The larger $n$, the softer the deformation. In many cases, existence of a first- or second-order isometry is sufficient to imply an infinite order isometry \cite{lewicka}.

Suppose we now consider the generalized class of deformations in which $\epsilon^I$ itself can vary with position on the shell's mid-surface. Computing the in-plane stresses induced by such a deformation to first order, one obtains
\begin{equation}\label{eq:strain1}
\gamma_{i j} \approx \sum_{I=1}^N \left( \partial_i \epsilon^I  T_j^I + \partial_j \epsilon^I  T_i^I \right),
\end{equation}
where $(T_i)^I \equiv \partial_{\epsilon^I} \mathbf{X} \cdot \partial_i \mathbf{X}$ evaluated at $\epsilon^I=0$. Thus, if we restrict ourselves to this class of deformations, the stretching content of the deformation can be made vanishingly small by making the gradient small. To this order, the total strain energy  can be condensed into a simple form,
\begin{equation}\label{eq:stretching}
E_{st} = \frac{t}{2} \int d^2x~\sqrt{\bar{g}} \sum_{I J} S^{i j}_{I J} \partial_i \epsilon^I \partial_j \epsilon^J,
\end{equation}
which must then be integrated over the area of the shell mid-surface. Notice, however, that the tensor $S^{i j}_{I J}$ can generically depend on position.

It is instructive to think of the example of an acoustic phonon in a crystal which, like other Nambu-Goldstone modes, is a gapless oscillation arising from a spontaneously-broken global symmetry. The broken translation symmetry of a crystal ensures that a long-wavelength oscillation of the atoms can have an arbitrarily low energy with increasing wavelength. Similarly, the ability to deform a shell along isometries without stretching energy cost ensures that the deformations resulting in Eq. (\ref{eq:strain1}) can also be made arbitrarily small. On the other hand, the stretching energy densities of non-isometric deformations scale with $Y t \gamma^2$. Thus, and in analogy with case of Nambu-Goldstone modes, I conjecture that  gapless, \textit{local} isometries dominate the diffuse, stretching deformations shells.

Unlike in the stretching energy, the bending energy will generically have terms depending on $\epsilon^I$, $\partial_i \epsilon^I$, and $\partial_i \partial_j \epsilon^I$ to quadratic order. However, because this paper is primarily concerned with very diffuse deformations and these terms are already down by $t^2$ compared to the stretching energy, I employ the approximation
\begin{equation}\label{eq:bending}
E_b \approx \frac{t^3}{24} \int d^2x~\sqrt{\bar{g}} ~\sum_{I J} B_{I J} \epsilon^I \epsilon^J,
\end{equation}
where $B_{I J}$ is a position-dependent bending matrix. These two expressions for $E_{st}$ and $E_b$ are the main result of this work.

One can also compute the higher-order corrections to $\gamma_{i j}$ for the Nambu-Goldston-like local isometries. These are
\begin{eqnarray}\label{eq:2ndorder}
\gamma_{i j}^{(2)} &=& \epsilon^J \partial_i \epsilon^I \partial_{\epsilon^I} \mathbf{X} \cdot \partial_j \partial_{\epsilon^J} \mathbf{X} + \epsilon^I \partial_j \epsilon^J \partial_{\epsilon^J} \mathbf{X} \cdot \partial_i \partial_{\epsilon^I} \mathbf{X}\\  
& & + \partial_i \epsilon^I \partial_j \epsilon^J \partial_{\epsilon^I} \mathbf{X} \cdot \partial_{\epsilon^J} \mathbf{X} + \epsilon^I \epsilon^J \partial_{\epsilon^I} \partial_{\epsilon^J} (\partial_i \mathbf{X} \cdot \partial_j \mathbf{X}).\nonumber
\end{eqnarray}
The last term of Eq. (\ref{eq:2ndorder}) vanishes for second order isometries but not necessarily in general. The first two terms can be compared directly to Eq. (\ref{eq:strain1}) and suggest that the limit of validity for the quadratic approximation used in Eq. (\ref{eq:stretching}) is, $\epsilon^J \partial_{\epsilon^J} \mathbf{X} \cdot \partial_i \partial_{\epsilon^I} \mathbf{X} \ll {T_i^I}$. Similarly, the third term of Eq. (\ref{eq:2ndorder}) yields the bound $\partial_i \epsilon^J \partial_{\epsilon^I} \mathbf{X} \cdot \partial_{\epsilon^J} \mathbf{X} \ll T_i^I$.

What remains is to find expressions for the first-order isometries. I require only that $0 = \partial_{\epsilon^I}|_0 g_{i j} \approx \partial_i \mathbf{X}_{0} \cdot \partial_j \partial_{\epsilon^I}|_0 \mathbf{X}_{\vec{\epsilon}} + \partial_j \mathbf{X}_{0} \cdot \partial_i \partial_{\epsilon^I}|_0 \mathbf{X}_{\vec{\epsilon}}$. Therefore, $D_j T_i^I +D_i T_j^I = 2 h^I h_{i j}$, where $D_i$ is the covariant derivative on the surface and $h^I = \mathbf{N} \cdot \partial_{\epsilon_I}|_0 \mathbf{X}$ is the component of the isometry normal to the surface. Since $h^I$ only appears algebraically, it can be eliminated to produce two equations in terms of the two unknown functions $T_i^I$.

I will focus on axisymmetric, closed shells, described by coordinates $(r,\theta)$, $\mathbf{X}(r,\theta) = \left( \rho(r) \cos \theta, \rho(r) \sin \theta,\int_{r_0}^{r} d\xi \sqrt{1-\rho'^2(\xi)} \right)$. In these coordinates, $r$ is the arc length along the symmetry axis and $\theta$ the azimuthal angle around the symmetry axis. The isometries can now be found in terms of a Fourier decomposition: $(T_i)^I = \sum_m (T_i)^{I m} e^{i m \theta}$. Assuming that the $\epsilon^I$ can only vary in $r$, I obtain the equations
\begin{eqnarray}\label{eq:isometry2}
\left( \frac{T_r}{\sqrt{1-\rho'^2}}\right)' &=& - i \frac{m}{\rho} \frac{\rho''}{(1-\rho'^2)^{3/2}} T_\theta\\
\left(\frac{T_\theta}{\rho^2} \right)' &=& -i \frac{m}{\rho^2} T_r\nonumber.
\end{eqnarray}
and
\begin{equation}\label{eq:h}
h = \frac{i m}{\rho \sqrt{1-\rho'^2} } T_\theta + T_r \frac{\rho'}{\sqrt{1-\rho'^2}},
\end{equation}
where the prime denotes a derivative with respect to $r$ and I have suppressed extraneous indices.

When $m=0$, Eqs. \ref{eq:isometry2} can be integrated in closed form: $T_r = \sqrt{1-\rho'^2(r)} C_1$, $T_\theta = \rho^2(r) C_2$, and $h= \rho'(r) C_1$, where $C_{1,2}$ are the two constants of integration. More generally, there is typically a two-parameter family of first-order isometries for each Fourier mode (though this need not always be the case \cite{audoly, audoly1}). It is useful to associate these constants in terms of more physical boundary conditions. Denoting $r=0$ as the circle at which a deformation will be applied, I choose $T_r(r=0)=\rho(0) \epsilon^{1,m=0}$ and $T_\theta(r=0) = 0$ for one family of isometries, and $T_r = 0$ and $T_\theta(r=0) = \rho(0)^2 \epsilon^{2,m=0}$ for the other. Thus,
\begin{eqnarray}
\partial_{\epsilon^{1, 0}} \mathbf{X} &=& \frac{1}{\sqrt{1-\rho'(0)^2}} \left[ \sqrt{1-\rho'(r)^2} \partial_r \mathbf{X} +  \rho'(r)\mathbf{N} \right]\nonumber\\  \label{eq:displacement}
\partial_{\epsilon^{2, 0}} \mathbf{X} &=& \frac{\rho^2(r)}{\rho^2(0)} \partial_\theta \mathbf{X}.
\end{eqnarray}
The first isometry controls translation along the symmetry axis of the shell. The second isometry controls rotations about the symmetry axis. Similarly, the $m=1$ modes describe translations transverse to and rotations of the symmetry axis of the shell. The $m=2$ mode on a cone is shown in Fig. \ref{fig:cone}.

The circle at $r=0$ plays a prominent role in Eq. (\ref{eq:displacement}) because of the choice of boundary conditions. For example, the $m=1$ isometries that rotate the shell do so about an axis perpendicular to the shell's symmetry axis and passing through the circle at $r=0$. As a consequence, the expressions for $S^{i j}_{I J}$ and $B_{I J}$ will depend explicitly on $r$, and would be superficially different were the boundary conditions applied elsewhere on the shell. However, the full symmetries of the shell energy under arbitrary rotations can be recovered by constructing linear combinations of isometries with different $m$. For example, combining a translation along the symmetry axis ($m=0$) followed by a rotation of the symmetry axis about the origin ($m=1$), one obtains an arbitrary infinitesimal rotation. Similarly, translations and rotations can be combined with higher order modes ($m \ge 2$) to obtain deformations of the shell anywhere along the symmetry axis. The choice of where to apply the boundary conditions is purely a convenience, because that is where we are imagining applying the deformation.

Though there is no general analytic solution to Eq. (\ref{eq:isometry2}), the same analysis can be performed for each Fourier mode. For notational simplicity, we separate the sum over $m$ and let the index $I$ be either $1$ or $2$. Then the strain components become
\begin{eqnarray}
\gamma_{r r} &=& \sum_{m I} 2 {\epsilon^{I m}}'(r) T_r^{I m} e^{i m \theta}\\
\gamma_{r \theta} &=& \sum_{m I} {\epsilon^{I m}}'(r) T_\theta^{I m} e^{i m \theta}\nonumber,
\end{eqnarray}
and $\gamma_{\theta \theta} = \mathcal{O}(\epsilon^2)$.

To see how these pieces fit together, I will first recover the case of a cylinder deformed at $r=0$ \cite{maha}, for which $\rho(r)=R$ is constant. There are two solutions for each $m$, which we normalize with either $T_r = R \epsilon^{1 m}$, $T_\theta=0$ or $T_r=0$,$T_\theta = R^2 \epsilon^{2 m}$. For the stretching energy, we need only
\begin{equation}
S^{r r}_{I J} = \mu R^2  \left(
\begin{tabular}{ccc}
$m^2 r^2/R^2 + \frac{\lambda+2 \mu}{\mu}$ & & $-i m r/R$\\
& & \\
$i m r/R$ & & $1$
\end{tabular}
\right)
\end{equation}
in Eq. (\ref{eq:stretching}). The bending energy is given by Eq. (\ref{eq:bending}) with
\begin{eqnarray}
B_{I J} &=& \frac{(\lambda+2 \mu) (m^2-1)^2}{R^2}\nonumber \\
&\times& \left(
\begin{tabular}{cc}
$4 \frac{\mu}{\lambda+2\mu} + m^2 r^2/R^2$ & $-i m^3 r/R$\\
$i m^3 r/R$ & $1$
\end{tabular}
\right).
\end{eqnarray}
These expressions are only valid for $m > 0$.

The Euler-Lagrange equations for the $(\epsilon^{1m}, \epsilon^{2m})$ are
\begin{equation}\label{eq:EL}
- \partial_r \left( S^{r r}_{I J} \partial_r \epsilon^{J m} \right) + \frac{t^2}{12} B_{I J} \epsilon^{J m} = 0,
\end{equation}
where repeated indices are summed. After some algebra, $\epsilon^{2 m}$ can be eliminated for $\epsilon^{1 m}$, leaving the single fourth order equation for $\epsilon^{1 m}$ 
\begin{equation}\label{eq:pinch}
\partial_r^4 \epsilon^{1 m} + \mathcal{O}(t^2) \partial_r^2 \epsilon^{1 m}+\frac{t^2 m^4 (m^2-1)^2}{12 R^6} \epsilon^{1 m} =0,
\end{equation}
where $\mathcal{O}(t^2)$ indicates a term of order $t^2$ that I will subsequently neglect. Setting $m=2$, Eq. (\ref{eq:pinch}) predicts a characteristic decay length $\lambda = (2 \sqrt{3})^{-1/2} t^{-1/2} R^{3/2}$. Denoting a persistence length $\ell_P = 2 \pi/\lambda$, as in Ref. \cite{maha}, one obtains  the correct scaling, the precession of the elliptical cross-section,  and a prefactor $\ell_P t^{1/2} R^{-3/2} = 2 \pi/\sqrt{2 \sqrt{3}} \approx 3.38$ that is not too far from the simulation value of, approximately, $4$ \cite{maha}. Another difference is that the independence of the prefactor on Poisson ratio. However, even in the small deformation analysis of Mahadevan \textit{et al.} \cite{maha}, the dependence on Poisson ratio is weak.

\begin{figure}[t]
\includegraphics[width=3.5in]{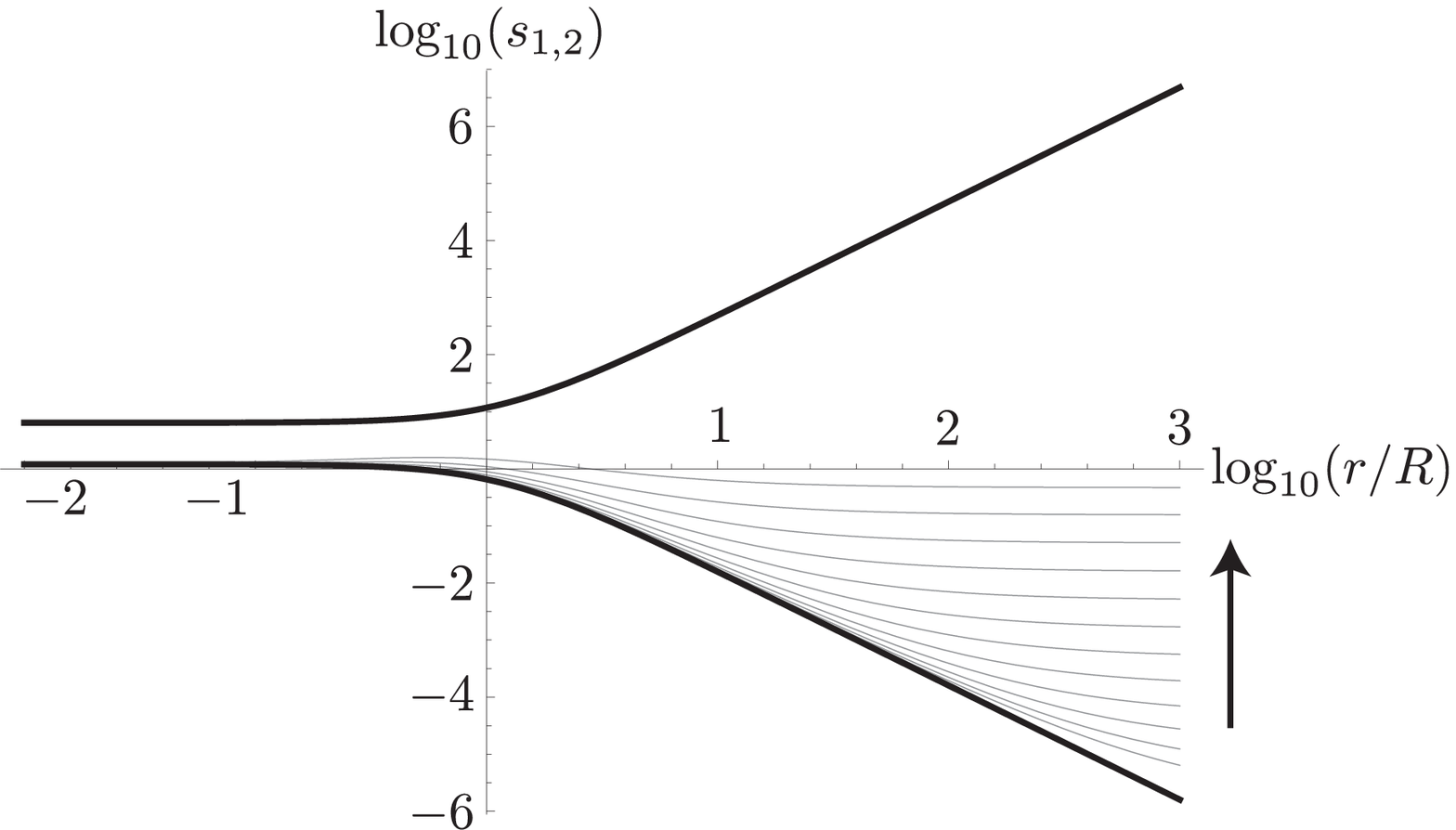}
\caption{Eigenvalues of the dimensionless stretching matrix for a ``pinch'' at $r=0$ on a cone [$\rho(r) = R + \alpha ~r$] with $\alpha = 10^{n-3}$ for $n=0-2.75$. Arrow points toward increasing $\alpha$. The result for a cylinder is shown as a thick line.}
\label{fig:se}
\end{figure}
\begin{figure}[t]
\includegraphics[width=3.5in]{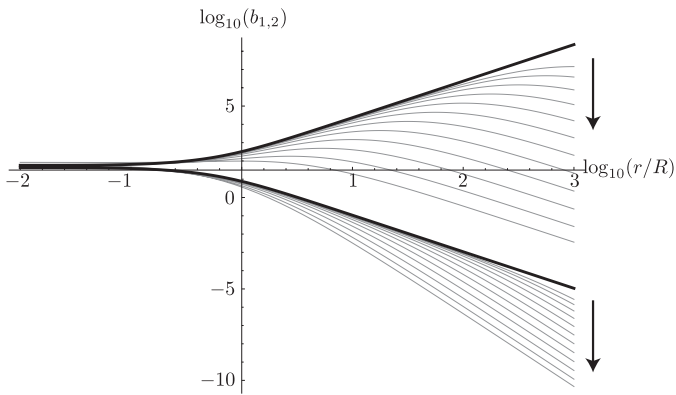}
\caption{Eigenvalues of the dimensionless bending matrix for a ``pinch'' at $r=0$ on a cone [$\rho(r) = R + \alpha ~r$] with $\alpha = 10^{n-3}$ for $n=0-2.75$. Arrows point toward increasing $\alpha$. The result for a cylinder is shown as a thick line.}
\label{fig:be}
\end{figure}
Eq. (\ref{eq:EL}) can be easily generalized to other axisymmetric shapes and solved numerically. The simplifications that occur for cylinders leading to Eq. (\ref{eq:pinch}) do not occur more generally, however. As a prototype of the more generic case, I consider a cone, for which $\rho = R + \alpha~ r$, deformed at one edge located at $r=0$. When $\alpha=0$, this shape recovers the cylinder previously discussed; when $\alpha=1$, however, the surface is a flat sheet with a circular hole of radius $R$ at its center. We first consider the two eigenvalues of $S^{rr}_{I J}/(\mu R^2)$ for $m=2$ and various $\alpha$, plotted in Fig. \ref{fig:se}, and the corresponding eigenvalues of $B_{I J} R^2/(\lambda+2 \mu)$, plotted in Fig. \ref{fig:be}. As $\alpha$ decreases, the eigenvalues approach those of the cylinder, $\alpha=0$, shown as thick lines.

\begin{figure}[t]
\includegraphics[width=3in]{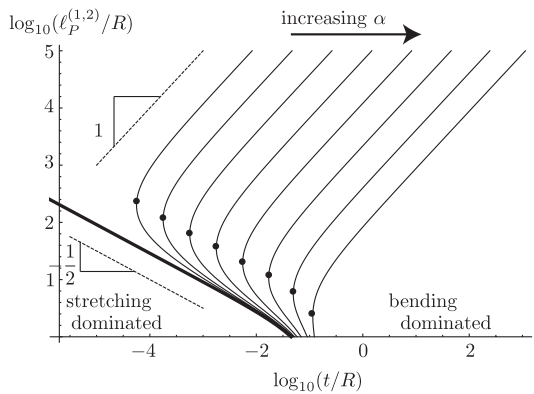}
\caption{Persistence length of a ``pinch'' at $r=R$ on a cone [$\rho(r) = R + \alpha~r$] with $\alpha = 10^{n-3}$ for $n=0-2.75$. The cylinder is shown as a thick line. The critical point at $t_c$ is denoted by a circle. The dashed lines have slopes $-1/2$ (bottom) and $+1$ (top) respectively.}
\label{fig:depth}
\end{figure}

Since the largest eigenvalue of the stretching matrix grows with increasing distance from the pinch, $r$, I will treat it as a constraint. Setting $\partial_r \epsilon^I$ to vanish along the largest eigenvector of $S_{I J}$, one obtains the approximation $m \partial_r \epsilon^{1} = \alpha i \partial_r \epsilon^2$ for $r \gg R$. Since the smallest bending eigenvalue is already down by a factor of $t^2/12$, it can be safely ignored. Consequently, only the smallest stretching and largest bending eigenmodes compete. For small $r$, the stretching eigenvalue dominates and the shell deforms along an isometry locally; at larger $r$, however, the largest bending eigenvalue may come to dominate and relax the deformation. Thus, I define the persistence length as the length at which the ratio of the smallest stretching and largest bending eigenvalues is one. For a general cone, there are two $r$ at which this occurs, which I denote $\ell_P^{(1)} < \ell_P^{(2)}$, both plotted in Fig. \ref{fig:depth} as a function of thickness for different values of $\alpha$. It is only in the singular limit $\alpha \rightarrow 0$, that a single length scale $\ell_P^{(1)} \sim t^{-1/2}$ emerges.

For $r < \ell_P^{(1)}$, the stretching energy dominates while for $\ell_P^{(1)} < r < \ell_P^{(2)}$, the bending energy dominates. For $r > \ell_P^{(2)}$, the stretching energy dominates again.
The multiple scales occurring in the cone have a natural interpretation: a deformation initially decays with a characteristic length scale $\ell_P^{(1)}$, but this decay halts beyond $\ell_P^{(2)}$ as the stretching energy comes to dominate again. The separation between these length scales vanishes at a critical thickness, $t_c$, below which the stretching energy always dominates. In this regime, diffuse deformations are essentially isometric. The critical thickness is not apparent on cylinders, since $t_c \rightarrow 0$ as $\alpha \rightarrow 0$.
On the other hand, when $\alpha$ approaches $1$, so that the cone becomes nearly flat, the normal component of the isometries diverges with $h^{I m} \sim \mathcal{O}[(1-\alpha^2)^{-1/2}]$. It is clear that both bounds of validity emerging from Eq. (\ref{eq:2ndorder}) will be violated and nonlinear terms in the strain, which couple modes with different $m$'s, will become relevant.

Finally, one can ask when one should expect a description of shell deformations in terms of local isometries to fail. As suggested by a simple scaling argument, the stretching energy gap of non-isometric deformations may still be sufficiently large that the Nambu-Goldstone-like local isometries dominate even when the quadratic approximation for the energy fails. That would suggest that the methods employed here are applicable quite generally to the deformations of shells. To know this for sure, however, one would have to consider the size of these gaps, a more complex calculation not undertaken here. It is tempting to conjecture that a shell will crumple or form a sharp feature precisely as these non-isometric modes become important. This will be the subject of a future investigation.

\begin{acknowledgements}
I am indebted to discussions with B.G. Chen and J.A. Hanna, and to E. Hohlfeld, L. Mahadevan, and P.M. Reis for comments on the manuscript. I acknowledge funding through NSF DMR-0846582. I wish to thank the Aspen Center for Physics, where this work was first conceived, for their hospitality.
\end{acknowledgements}

\end{document}